\documentclass[a4paper,twocolumn,%tightenlines,
english,aps,pre,floatfix,showpacs]{revtex4}
\usepackage[T1]{fontenc}
\usepackage[latin1]{inputenc}
\usepackage{amsmath}
\usepackage{babel}
\usepackage{graphics}
\usepackage{amssymb}


\begin{document}
\title
{Distribution of local Lyapunov exponents in spin-glass dynamics}
\author {S.L.A. \surname{de Queiroz}}

\email{sldq@if.ufrj.br}

\affiliation{Instituto de F\'\i sica, Universidade Federal do
Rio de Janeiro, Caixa Postal 68528, 21941-972
Rio de Janeiro RJ, Brazil}

\author {R. B. \surname{Stinchcombe}}

\email{stinch@thphys.ox.ac.uk}

\affiliation{Rudolf Peierls Centre for Theoretical Physics, University of
Oxford, 1 Keble Road, Oxford OX1 3NP, United Kingdom}

\date{\today}

\begin{abstract}
We investigate the statistical properties of local Lyapunov exponents
which characterize magnon localization in the one-dimensional 
Heisenberg-Mattis spin glass (HMSG) at zero temperature, by means
of a connection to a suitable version of the Fokker-Planck (F-P) equation. We consider 
the local Lyapunov exponents (LLE), in particular the case of {\em instantaneous} LLE.
We establish a connection between the transfer-matrix recursion relation for 
the problem, and an F-P equation governing the evolution of the probability
distribution of the instantaneous LLE. The closed-form 
(stationary) solutions to the F-P equation are in excellent accord with 
numerical simulations, for both the unmagnetized and magnetized versions of the
HMSG. Scaling properties for non-stationary conditions are derived from  the F-P 
equation in a special limit (in which diffusive effects tend to vanish), and also shown 
to provide a close description to the corresponding numerical-simulation data.     
\end{abstract}
%05.10.Gg - Stochastic analysis methods (Fokker-Planck, Langevin, etc.)
%75.10.Nr - Spin-glass and other random models
%63.20.Pw - Localized modes
\maketitle
%\tightenlines
 
\section{Introduction} 
\label{intro}
The analytic treatment of quenched disordered systems in
Condensed Matter Physics invokes many concepts from statistical theory. 
Among these, we shall be concerned in this paper with the connection between
the probability distribution of Lyapunov 
characteristic exponents for the (equilibrium) problem of low-lying magnetic 
excitations in spin glasses, and the Fokker-Planck (F-P) equation~\cite{fp1}, which is a 
key element in the description of non-equilibrium stochastic processes. 

Links between an F-P equation and stationary distributions of
physical quantities in a transfer-matrix (TM) description have been established, usually
in the context of calculating the probability distribution functions (PDF) of electronic 
conductance, in one-dimensional (1d) or quasi-1d
noninteracting electronic systems with disorder 
(Anderson localization)~\cite{ond82,ond83,mpk88,mc91,end96,been97,bmsa98,bfmr06}.
Specifically, one considers the distribution of the TM itself (or, equivalently,
its transmission eigenvalues), between the left and right extremes of a wire of length 
$L \gg 1$ lattice spacings. The F-P equation is set up to account for the infinitesimal
changes in the TM, caused by a small length variation $\delta L$. The quantities whose  
PDF is calculated are, therefore, aggregate in the sense that they explicitly
incorporate all contributions for the TM, say, from $x=0$ to $L$, i.e., they represent
{\em global} Lyapunov exponents. This is in contrast with the approach we take here; 
as shown below, we shall concentrate on the PDF of {\em local} Lyapunov exponents, which
appear not to have been as thoroughly exploited as their global counterparts (at least
in the Condensed-Matter context of localization and similar problems). 

As a model system to apply the ideas developed here, we consider spin waves 
(magnons) in a simplified spin glass model, the so-called 
Heisenberg-Mattis spin glass (HMSG)~\cite{dcm76}. 
In 1d, the dynamics of HMSGs turns out to exhibit many non-trivial 
features~\cite{ds77,clh77,ds79,ch79,sp88,ps88,ga04,dqrbs06}, including dynamic 
exponents~\cite{sp88,ps88,dqrbs06} different from the standard hydrodynamic 
predictions~\cite{hs77}. 

In 1d, all
eigenstates are localized for any amount of disorder, though as 
energy  $\omega \to 0$, the  localization length $\lambda(\omega)$
diverges continuously. This, in turn, suggests the applicability of scaling
concepts in the low-frequency, long-wavelength limit.

Operationally, one can obtain $\lambda(\omega)$ in one-dimensional systems by
using a TM approach, and extracting the smallest Lyapunov
exponent (whose inverse corresponds to the largest localization length) which
arises from repeated iteration of the TM. This has been done for
HMSG chains~\cite{bh89,ew92,ah93}, and can be extended to HMSG in $d=2$ and 
$3$~\cite{dqrbs06}.  

Lyapunov exponents are well defined quantities 
in the sense that, for $N \gg 1$ iterations of the TM, the width of the distribution 
of their corresponding estimates shrinks to zero~\cite{ranmat} (as
$N^{-1/2}$ in many cases of interest, as a consequence of the central limit
theorem). However, it has been shown~\cite{gbp88,abk91,ab93,pr99,dr03} that the so-called 
{\em local} Lyapunov  exponents (LLE, to be defined more accurately below) exhibit 
non-trivial distributions whose width remains finite even as $N$ becomes fairly large
(compared, e.g., with the localization length). Furthermore, LLEs may provide
a wealth of information specific to the dynamics of the associated physical system,
which does not manifest itself in the aggregated, $\delta$-function like behavior of
their global counterparts.

In this paper, we investigate the statistical properties of
local Lyapunov exponents in the 1d HMSG at zero temperature, by means
of a connection to a suitable version of the F-P equation. 
  
In Section~\ref{sec:2} we recall pertinent aspects of the HMSG, 
specializing to 1d. In Section~\ref{sec:3}, we first consider the aggregate effects 
which characterize the (global) Lyapunov exponents for the system under study; then we
investigate local exponents (LLE), in particular the case of {\em instantaneous} LLE.
We establish a connection between the TM-iterated recursion relation for the
problem, and a suitable F-P equation governing the evolution of the probability
distribution of the instantaneous LLE. We show that the closed-form 
(stationary) solutions to the F-P equation are in excellent accord with 
numerical simulations, for both the unmagnetized and magnetized versions of the 1d 
HMSG. Scaling properties for non-stationary conditions are derived from  the F-P 
equation in a special limit (in which diffusive effects tend to vanish), and also shown 
to provide a close description to the corresponding numerical-simulation data.     
Finally, in section~\ref{sec:conc}, concluding remarks are made. 

\section
{Heisenberg-Mattis spin glasses: scaling in 1d}
\label{sec:2}

We consider Heisenberg spins on sites of a hypercubic 
lattice, with nearest-neighbor couplings:
\begin{equation}
{\cal H}= -\sum_{\langle i,j \rangle} J_{ij}\,{\mathbf S}_i \cdot
{\mathbf S}_j
\label{eq:1}
\end{equation}
The bonds are randomly taken from a quenched, binary probability
distribution,
\begin{equation}
P(J_{ij})= p\,\delta (J_{ij}-J_0)+ (1-p)\,\delta (J_{ij}+J_0)\ ;
\label{eq:2}
\end{equation}
here we shall mainly consider $p=1/2$ (unmagnetized spin glass).
 
The Mattis model ascribes disorder to sites rather
than bonds ($J_{ij} \to J_0\,\zeta_i\,\zeta_j$), so that the Hamiltonian
reads:
\begin{equation}
{\cal H}_M= -J_0\sum_{\langle i,j \rangle} \zeta_i\,\zeta_j\,{\mathbf S}_i
\cdot {\mathbf S}_j\ ,
\label{eq:3}
\end{equation}
where $\zeta_i=+1,\ -1$ with probability $p$, ($1-p$). Then $p=1/2$ is the
Mattis spin glass while $p \neq 1/2$ corresponds to a magnetized Mattis model.
This way, the overall energy is minimized by making $S_i^z =\zeta_i\,S$,
which constitutes a (classical) ground state of the Hamiltonian
Eq.~(\ref{eq:3}). Consideration of the spin-wave equations of motion
(see, e.g., Refs.~\onlinecite{sp88,ps88,dqrbs06}) gives, with $\hbar=1$:
\begin{equation}
i\,\zeta_i\,d u_i/dt = \sum_j J_0\,\left(u_i -u_j\right)\ .
\label{eq:4}
\end{equation}
where the $u_i$ are Mattis-transformed local (on-site) spin-wave
amplitudes, and the sum is over sites $j$ which are nearest neighbors of $i$.
For the eigenmodes with frequency $\omega$ (in units of the exchange
constant $J_0$), Eq.~(\ref{eq:4}) leads to
\begin{equation}
\omega\,\zeta_i\,u_i=  \sum_j \left(u_i -u_j\right)\ .
\label{eq:5}
\end{equation}
The relationship of frequency to wave number, $k$ (in the context of
localization, this corresponds to $\lambda^{-1}$), at low energies is
characterized by the dynamic exponent $z$:
\begin{equation}
\omega \propto k^z\ .
\label{eq:6}
\end{equation}
For $p = 1/2$ in 1d, it was predicted  analytically~\cite{sp88,ps88}, and 
verified by numerical calculations~\cite{bh89,ew92,ah93}, that $z=3/2$.
Still in 1d, but at general $p$, Eq.~(\ref{eq:5}) becomes
\begin{equation}
(2-\zeta_i\,\omega)\,u_i = u_{i-1}+u_{i+1}\ .
\label{eq:1d}
\end{equation}
A TM approach\cite{ps81,hori} can be formulated,
giving~\cite{sp88,ps88,bh89,ew92}:
\begin{equation}
\begin{pmatrix}{u_{i+1}}\cr{u_i}\end{pmatrix}=
\begin{pmatrix}{2-\zeta_i\,\omega}&{-1}\cr{1}&{0}\end{pmatrix}
\begin{pmatrix}{u_i}\cr{u_{i-1}}\end{pmatrix} \equiv T_i(\omega)\,
\begin{pmatrix}{u_i}\cr{u_{i-1}}\end{pmatrix}\ . 
\label{eq:1dtm}
\end{equation}
The procedure for calculating Lyapunov exponents in this case
is the same as that used for Anderson localization problems~\cite{ps81}.
Indeed, in both cases the TM is symplectic, and one can use Oseledec's
theorem and dynamic filtration~\cite{ranmat} to extract the smallest 
Lyapunov exponent, whose inverse is the largest localization length. 

\section{Lyapunov exponents}
\label{sec:3}

\subsection{Aggregate effects: the $N\to \infty$ limit}
\label{sec:3a}

The Lyapunov exponent emerging from iteration of Eq.~(\ref{eq:1dtm}) is given by
\begin{equation}\gamma (\omega)=\lim_{N\rightarrow \infty}\frac{1}{N}\ln \left\Vert 
\left( 
\prod_{i=1}^{N}T_{i}(\omega)\right)\vert v_{0}\rangle \right\Vert\,,
\label{eq:3.1}
\end{equation}
where $ \vert v_{0}\rangle \equiv\begin{pmatrix}{u_1}\cr{u_0}\end{pmatrix}$ 
is an arbitrary initial vector of unit modulus.

The LLE, $\gamma (N,\omega)$, is defined~\cite{gbp88,abk91,ab93,pr99,dr03} 
as the finite-$N$ version of Eq.~(\ref{eq:3.1}). For the sake of
completeness, before going further we exhibit the evolution of the statistics of 
$\gamma (N,\omega)$ (hereafter referred to as $\gamma_N$ for short) against increasing 
$N$. We consider the TM given in
Eq.~(\ref{eq:1dtm}), with $p=1/2$ and $\omega=0.015$ (fixed), for which case existing 
numerical results~\cite{bh89,ew92,dqrbs06} give $\lambda \approx 57$ lattice spacings,
i.e., $\gamma \approx 0.0175$.
\begin{figure}
{\centering \resizebox*{3.3in}{!}{\includegraphics*{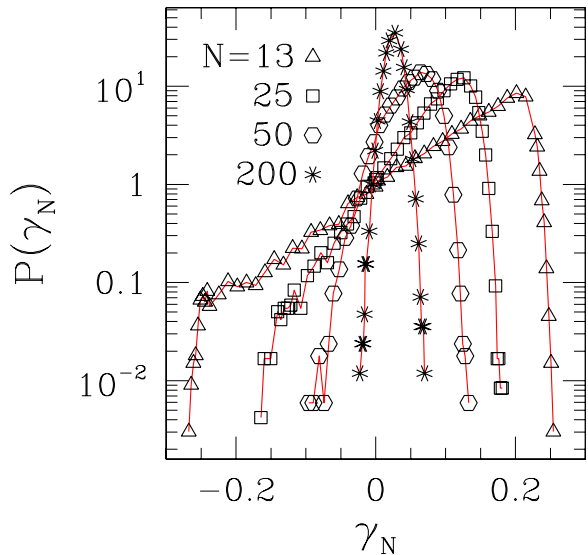}}}
\caption{(Color online) Normalized histograms of occurrence of LLE $\gamma_N$ for $p=1/2$ 
(spin glass), for several values of $N$. For each histogram, 
$N_s=10^5$ samples were taken. Here, $\omega=0.015$.
} 
\label{fig:lleb}
\end{figure}
\begin{figure}
{\centering \resizebox*{3.3in}{!}{\includegraphics*{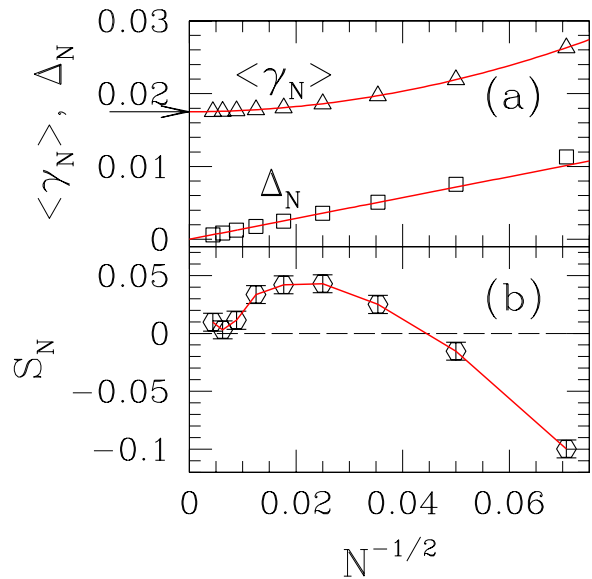}}}
\caption{(Color online) Estimates from the statistics of LLE $\gamma_N$ data, for
$N =2^k\times 100,\ k=1,\cdots,9$, against $N^{-1/2}$. For each $N$, 
$N_s=10^5$ samples were taken. Here, $p=1/2$, $\omega=0.015$. 
(a) Mean values $\langle \gamma_N\rangle$ and rms deviations $\Delta_N$.
Arrow on vertical axis
indicates estimate from a single sample, with $N=10^7$. Full lines 
are, respectively, quadratic ($\langle \gamma_N\rangle$) and linear 
($\Delta_N$) fits of data, the latter taking only $N \geq 800$
into account. (b) Skewness $S_N$. Error bars correspond to 
incompletely-sampled Gaussian distributions with $N_s=10^5$  (see text).
} 
\label{fig:llen}
\end{figure}

In Fig.~\ref{fig:lleb}, one sees that the probability distribution function (PDF),
$P(\gamma_N)$, of $\gamma_N$ takes on an approximately Gaussian shape only for
$N \gtrsim \lambda$~. Fig.~\ref{fig:llen} (a) provides a quantitative check on
the increasing narrowness of the PDF for $N \gg \lambda$, and on the convergence
process $\langle\gamma_N\rangle \to \gamma$. One can infer the accuracy of the 
single-sample 
estimate corresponding to $N=10^7$ (denoted by an arrow pointing to the vertical axis in 
Fig.~\ref{fig:llen} (a)), by extrapolating the $N$-dependence
of the width $\Delta_N$ of the PDF against $N \leq 51200$ (shown in the main
diagram). The result is $\gamma=0.01752(4)$, hence the associated error bar would be 
invisible on the scale of the Figure.  

The simplest quantitative indicator of whether the PDF actually turns Gaussian with 
increasing $N$ [in the process of becoming a $\delta -$ function, characteristic of the 
self-averaging property just demonstrated] is 
its skewness~\cite{nr}, $S_N \equiv \langle 
((\gamma_N-\langle\gamma_N\rangle)/\Delta_N)^3\rangle$. One must
bear in mind that, because the number of samples $N_s$ is finite (incomplete sampling), 
$S_N$ itself will have a distribution. While this is true also for 
$\langle\gamma_N\rangle$ and $\Delta_N$, the effects on $S_N$ are much more prominent
since it is the ratio of two quantities which, in the present case, both vanish as $N, 
N_s \to \infty$. Indeed, we have checked that, for the data shown in 
Fig.~\ref{fig:llen} (a), corresponding to $N_s=10^5$, estimates of 
$\langle\gamma_N\rangle$ and $\Delta_N$ from distinct sequences of pseudo-random
numbers usually differ only by a few parts in $10^4$. On the other hand, it is 
known~\cite{nr} that, for an incompletely sampled Gaussian distribution,
the width of its own skewness distribution is approximately $\sqrt{6/N_s}$ ($\approx 
0.0078$ here). Since our own large-$N$ PDFs are, to zeroth-order, Gaussian, we shall use 
this value as a lower bound to the uncertainty of our estimates for $S_N$. The results 
are displayed in  Fig.~\ref{fig:llen} (b). Given the trend exhibited by the calculated 
$S_N$ for the three largest values of $N$ used (especially in contrast with the
smaller-$N$ region), it appears safe to conclude that
the skewness is in fact approaching zero, within the pertinent error bars.
Thus, the overall evidence is compatible with a limiting Gaussian form for the PDF
of the LLEs as $N \to \infty$. 

\subsection{Local effects and the Fokker-Planck equation}
\label{sec:3b}

We now turn to the opposite limit, which is our main concern here. Instead of 
considering the aggregate effect of contributions to $\gamma_N$, we shall 
analyse the PDFs of such contributions separately.

The case $N=1$ of Eq.~(\ref{eq:3.1}) is often denoted as ``instantaneous'' 
LLE~\cite{dr03}. Here, we shall use the term in the following way. The
{\it instantaneous LLE at} $i=M$, to be denoted by $s(M)$, is the local
contribution to the LLE, given at $i=M$, of the multiplying process denoted in
Eq.~(\ref{eq:3.1}). One then has:
\begin{equation}
\gamma_N = \frac{1}{N} \sum_{M=1}^{N}s(M)\ .
\label{eq:instdef}
\end{equation}   
For a one-dimensional mapping, the instantaneous LLE, as defined above, 
would be simply the local stretch factor~\cite{dr03}.
In the present case, where even in 1d the TM is $2 \times 2$, 
it is known that the eigenvectors (Lyapunov basis) of a product of random
matrices are only {\em local} properties, as opposed to the corresponding eigenvalues
(i.e., the Lyapunov spectrum) which are global ones~\cite{ranmat}.
Thus, one has a rotation of the local Lyapunov basis as the TM is repeatedly iterated.
As the TM is symplectic, in 1d this means that the (two) local 
eigenvalues are inverse of each other, so $s(M)$ can be extracted by 
suitable analysis of the growth factor associated to a given member of the
Lyapunov basis. 
 
The connection with the F-P equation proceeds as follows.
One gets, by taking the continuum limit of Eq.~(\ref{eq:1d}):
\begin{equation}
\omega\,\zeta(\ell)\,u(\ell)=\frac{\partial^2 u(\ell)}{\partial \ell^2}\ ,
\label{eq:cont1}
\end{equation}
where $\ell$ stands for position along the axis. One sees that
$u(\ell)$ is the result of accumulated contributions from the instantaneous Lyapunov
exponent $s(\ell^{\,\prime})$, $0 \leq \ell^{\,\prime} \leq \ell$:
\begin{equation}
u(\ell) =\exp \left(\int_0^\ell s(\ell^{\,\prime})\,d\ell^{\,\prime}\right)\ .
\label{eq:sdef}
\end{equation}
Therefore, the right-hand side of Eq.~(\ref{eq:cont1}) turns into: 
\begin{equation}
\frac{\partial^2 u(\ell)}{\partial \ell^2}=\left(s^2(\ell)+\frac{\partial  
s(\ell)}{\partial \ell}\right)\, u(\ell)\ .
\label{eq:svsu}
\end{equation}
Eq.~(\ref{eq:cont1}) then yields the following
equation for the stochastic variable $s(\ell)$:
\begin{equation}
s^2(\ell)+\frac{\partial s(\ell)}{\partial \ell} =
\omega\,\zeta(\ell) \ .
\label{eq:stoch}
\end{equation}
In the general case $\zeta$ is a binary-distributed variable
with mean ${\cal M}_1=2p-1$ and variance ${\cal M}_2=4p(1-p)$.
The F-P equation corresponding to the evolution of the probability 
distribution of $s$ is, 
with $\ell$ naturally assuming the role of time-like variable~\cite{fp1}:
\begin{eqnarray}
\frac{\partial P(s,\ell)}{\partial \ell}= 
\frac{\partial}{\partial s}\left((s^2+\omega\,{\cal M}_1)\,
P(s,\ell)\right)+
\nonumber\\
 +\frac{1}{2}\frac{\partial^2}{\partial 
s^2}\left({\cal M}_2\,\omega^2\,P(s,\ell)\right)\ .
\label{eq:fp-mag}
\end{eqnarray}

\subsection{The unmagnetized HMSG}
\label{sec:3c}

In this subsection we take the special case of an unmagnetized HMSG 
($p=1/2$). Eq.~(\ref{eq:fp-mag}) then becomes:
\begin{equation} 
\frac{\partial P(s,\ell)}{\partial \ell}=\frac{\partial}{\partial s}\left(s^2\,
P(s,\ell)\right)+\frac{1}{2}\frac{\partial^2}{\partial 
s^2}\left(\omega^2\,P(s,\ell)\right)\ .
\label{eq:fp1}
\end{equation}
Assuming stationarity, $P(s,\ell)\to  P(s)$ (which corresponds to the regime $\ell 
\gtrsim \lambda$), the equation to solve is:
\begin{equation}
\frac{\partial}{\partial s}\left(s^2\,
P(s)\right)+\frac{1}{2}\frac{\partial^2}{\partial s^2}\left(\omega^2\,P(s)\right)=0\ .
\label{eq:fp2}
\end{equation}
Thus, one gets:
\begin{equation}
P(s) ={\tilde C}\,\exp\left(-2s^3/3\omega^2\right)\,\int_{-\infty}^s 
\exp\left(2y^3/3\omega^2\right)\,dy\ .
\label{eq:fp3}
\end{equation}
It is immediate to see that the $3/2$ power, referred to in connection 
with Eq.~(\ref{eq:6}), arises in the scaling variable. Furthermore, it is
clear from the derivation of the present results that they apply only in the scaling 
regime, i.e., $\omega \to 0$, small $|s|$, $\ell \gtrsim \lambda$.

With $\int_{-\infty}^\infty dy\,\int_{-\infty}^y dx\,\exp(x^3-y^3) \equiv C_0
\approx 4.8$, the normalization of Eq.~(\ref{eq:fp3}) implies 
\begin{equation}
{\tilde C}={\tilde C}(\omega)=\frac{1}{(3\omega^2/2)^{1/3}\,C_0}\ .
\label{eq:norm}
\end{equation}
 
\begin{figure}
{\centering \resizebox*{3.3in}{!}{\includegraphics*{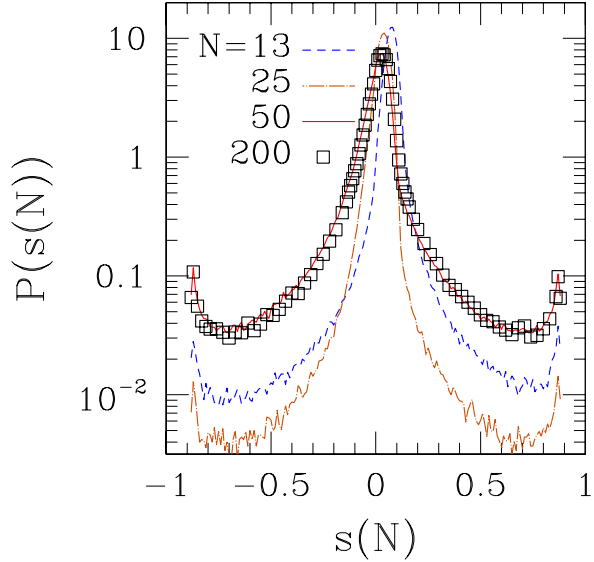}}}
\caption{(Color online) Normalized histograms of occurrence of instantaneous LLE $s(N)$ 
for several values of $N$. For each histogram, 
$N_s=10^6$ samples were taken. Here, $p=1/2$, $\omega=0.015$.
} 
\label{fig:inst}
\end{figure}

The numerical verification of the ideas just presented begins by checking the behavior
of the PDFs for $s(N)$, against varying $N$. Selected data are shown in 
Fig.~\ref{fig:inst}.
One can see that, for the ranges of $N$ considered in the Figure, the overall shape of 
the PDFs remains roughly constant, though a slight narrowing and shifting of the central
peak take place as $N$ increases. This is in contrast with the LLEs exhibited
in Fig.~\ref{fig:lleb}, for which variation of $N$ in the same range was accompanied by
a pronounced change in shape and width of the respective PDFs. For $1\leq N \lesssim 13$
[not shown in the Figure], the PDF  for $s(N)$ starts nearly flat at $N=1$, and gradually 
develops both the central peak, and
the small ``wings'' at the ends (whose relevance in Fig.~\ref{fig:inst} is overemphasized 
by the logarithmic scale on the vertical axis). The very good superposition of the 
$N=50$ and $200$ data in the Figure indicates the predicted trend towards a fixed, 
$N$-independent  form, which corresponds to stationarity of the F-P equation 
[see Eq.~(\ref{eq:fp2})]. This has 
been confirmed by examination of numerical PDFs for $100 \leq N \leq 600$.
Recalling that the localization length for $\omega=0.015$, as is the case of  
Fig.~\ref{fig:inst}, is $\lambda \approx 57$ lattice spacings, we see that the
condition for an $N$-independent PDF is indeed $N \gtrsim \lambda$.

\begin{figure}
{\centering \resizebox*{3.3in}{!}{\includegraphics*{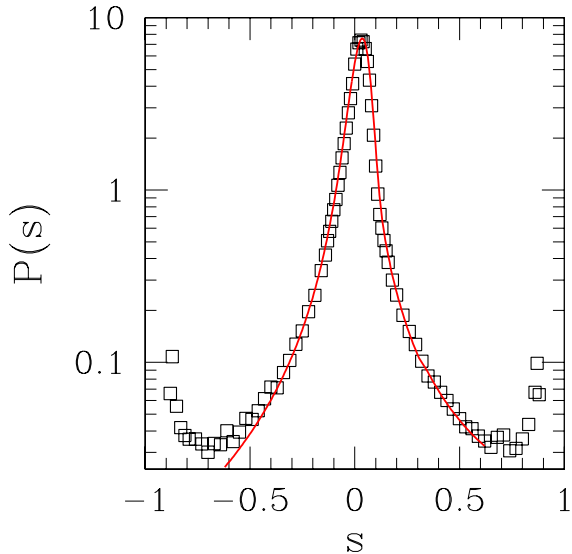}}}
\caption{(Color online) Squares: normalized histogram of occurrence of instantaneous LLE 
$s$, from $N_s=10^6$ samples, for $N=200$, $p=1/2$, $\omega=0.015$. Continuous line is the
analytical form, Eq.~({\protect{\ref{eq:fp3}}}), normalized and with appropriate
scaling of variables [see also Eq.~({\protect{\ref{eq:norm}}})].  
} 
\label{fig:exfit}
\end{figure}

We now test whether the form given in Eq.~(\ref{eq:fp3}) describes our numerical data.
Figure~\ref{fig:exfit} shows that the agreement is excellent, in the region 
$-0.5 \lesssim s \lesssim 0.5$, which is where scaling is expected to hold.
Note that there are no adjustable parameters in the fit, all scale factors being
provided by normalization and scaling considerations.
\begin{figure}
{\centering \resizebox*{3.3in}{!}{\includegraphics*{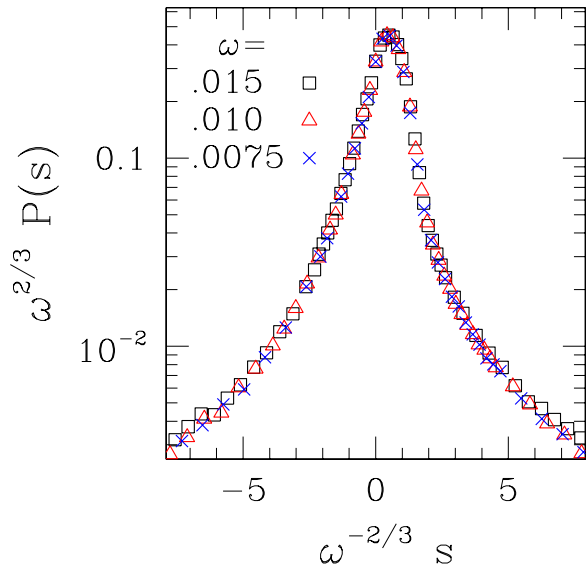}}}
\caption{(Color online) Scaling plots of normalized histograms of occurrence of 
instantaneous LLE , as suggested by  Eq.~({\protect{\ref{eq:fp3}}}), for
assorted energies. In all cases, $p=1/2$, $N_s=10^6$ samples, $N=200$ (so the condition 
$N \gtrsim \lambda(\omega)$ is always obeyed).
} 
\label{fig:scale}
\end{figure}

In Fig.~\ref{fig:scale} we test for consistency of scaling among numerical data 
for different energies~$\omega$, as suggested by Eq.~(\ref{eq:fp3}).
Again, the agreement with predictions is excellent, provided that the conditions
upon which Eq.~(\ref{eq:fp3}) was derived are obeyed. 

\subsection{The magnetized HMSG ($p \neq \frac{1}{2}$)}
\label{sec:3d}

For the magnetized case $p \neq 1/2$ we return to the general form of the F-P equation, 
Eq.~(\ref{eq:fp-mag}), whose stationary solution is given by:
\begin{equation}
P(s) ={\tilde C}^{\,\prime}\,\exp\left(-f(s)\right)\,\int_{-\infty}^s
\exp\left(f(y)\right)\,dy\ ,
\label{eq:fp-mag2}
\end{equation}
where
\begin{equation}
f(y)=f(y,\omega,p)=\frac{1}{\omega^2{\cal M}_2}\,\left(\frac{2y^3}{3}+2\omega{\cal 
M}_1\,y\right)\ .
\label{eq:fp-mag3}
\end{equation}
In analogy with Sec.~\ref{sec:3c}, we first check whether Eq.~(\ref{eq:fp-mag2}) gives a 
faithful description of simulational data, for given $p$ and $\omega$.
\begin{figure}
{\centering \resizebox*{3.3in}{!}{\includegraphics*{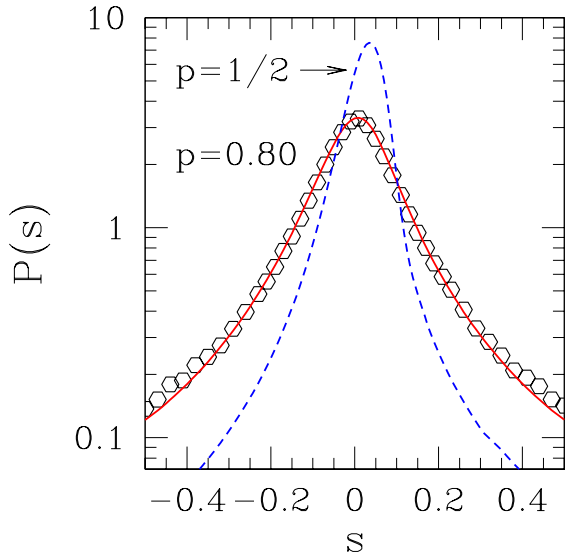}}}
\caption{(Color online) Hexagons: normalized histogram of occurrence of instantaneous LLE 
$s$, from  $N_s=10^6$ samples, for $\omega=0.015$, $p=0.80$ (here we used $N=600$, as the 
localization length is $\lambda \approx 498$). Continuous line is the
analytical form, Eq.~({\protect{\ref{eq:fp-mag2}}}), normalized. The dashed line is 
the analytical result for $p=1/2$, from Eq.~({\protect{\ref{eq:fp3}}}).  
} 
\label{fig:exfitp8}
\end{figure}
In Figure~\ref{fig:exfitp8}, calculated data for $p=0.8$, $\omega=0.015$ are shown,
together with the analytical result from Eq.~(\ref{eq:fp-mag2}) for the same values of
$p$ and $\omega$. Apart from an overall normalization factor
which affects only the vertical scale, there are no adjustable parameters. Again, the 
agreement is very good. 

In order to emphasize the degree to which the shape and range 
of the PDFs are affected by variations in the physical parameters, 
Figure~\ref{fig:exfitp8} also shows
the PDF for the unmagnetized case, at the same energy $\omega=0.015$. As $p$ increases 
from $1/2$, the general trend is towards reduction of the skewness (which is $\approx 
-0.32$ at $p=1/2$, $-0.02$ at $p=0.80$, and [in absolute value] $< 0.01$ at $p=0.95$). 
The peak of the distribution, which is about one standard deviation away from the origin 
at $p=1/2$, approaches $s=0$ as $p$ approaches unity. This process is accompanied by a
broadening  of the PDF. 

We now proceed to framing the preceding observations within a crossover description.
One sees that, because $f(y)$ in Eq.~(\ref{eq:fp-mag3}) is a 
polynomial in $y$, for $p \neq 1/2$ it is not possible to develop scaling arguments
(over the whole range of the variable $s$) 
similar to those invoked for $p=1/2$, and graphically depicted in Fig.~\ref{fig:scale}. 
Nevertheless, analysis of Eq.~(\ref{eq:fp-mag3}) shows that the crossover 
away from unbiased ($p=1/2$) behavior is governed by the dimensionless ratio
$z \equiv (\omega\,{\cal M}_1)^{1/2}/(\omega^2\,{\cal M}_2)^{1/3}$ ($\approx 
[(p-1/2)/\omega^{1/3}]^{1/2}$ away from $p \to 1$). This way,
$s$ scales with $\omega^{2/3}$ for $z \ll 1$, and with $\omega/(p-1/2)$
for $z \gg 1$. 
Analysis of the asymptotic behavior of $P(s)$, as given by Eqs.~(\ref{eq:fp-mag2})
and~(\ref{eq:fp-mag3}), shows that, with $z$ defined as above and $v\equiv 
s\,\omega^{-2/3}$, one has for the tails of the distribution:
\begin{equation}
R(v) \sim \frac{C}{v^2+z^2}\, \quad |v| \gg 1\ ,
\label{eq:asymp1}
\end{equation}
where $R(v) \equiv \omega^{2/3}\,P(s)$, and (for $z \gg 1$) the normalization constant $C 
\propto z$. 

Numerical checks of Eq.~(\ref{eq:asymp1}) against simulational data must be carried out
for suitable ranges of $p$, $\omega$, $s$ (equivalently, $z$, $v$), such that
(i) asymptotic behavior has already set in (i.e. large $|s|$, $|v|$), and (ii) one is 
still within the scaling regime (which implies small $|s|$). Thus one expects
Eq.~(\ref{eq:asymp1}) to hold only within windows of varying width, which may or may not
be easy to identify against statistical noise.

\begin{figure}
{\centering \resizebox*{3.3in}{!}{\includegraphics*{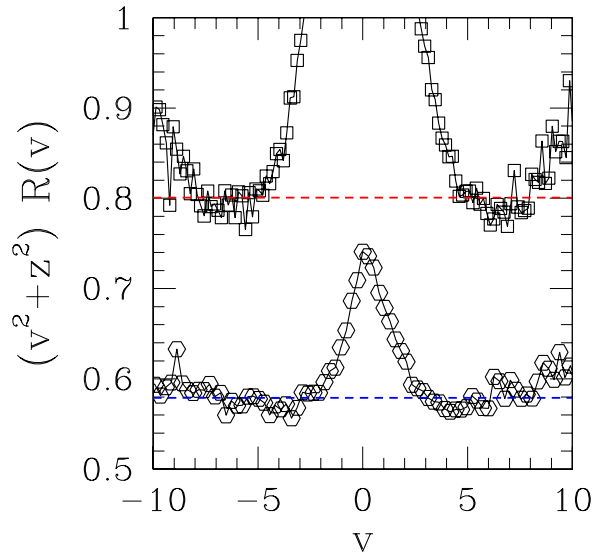}}}
\caption{(Color online) Symbols are data from normalized histograms of occurrence of 
instantaneous LLE  $s$, from  $N_s=10^6$ samples, plotted as suggested by
Eq.~({\protect{\ref{eq:asymp1}}}). See text for definitions of $z$, $v$, $R(v)$.
Hexagons: $p=0.8$, $\omega=0.0075$ ($z \approx 2.03$);
squares: $p=0.95$, $\omega=0.015$ ($z \approx 3.32$). Horizontal lines 
mark central estimates of averages of $(v^2+z^2)\,R(v)$, respectively along  
$-9 < v < -3$ (lower) and $-8 < v < -4$ (upper).
} 
\label{fig:asymp}
\end{figure}

In Figure~\ref{fig:asymp} we show that it is indeed possible to find intervals of 
$v$ along which a plot of $(v^2+z^2)\,R(v)$ is rather flat, as
suggested by Eq.~(\ref{eq:asymp1}). While the evidence is somewhat smeared
for the  upper set of data (corresponding to $p=0.95$, $\omega=0.015$, i.e. $z_1 \approx 
3.32$), it is clear for the lower one (for which $p=0.80$, $\omega=0.0075$, i.e. $z_2 
\approx 2.03$). Furthermore, the ratio of the averages of $(v^2+z^2)\,R(v)$
 (each denoted by a horizontal line 
in the Figure) along the respective flat sections [these latter defined with an	
inevitable degree of arbitrariness] is $\approx 1.4(1)$. This compares reasonably well 
with the
predicted value $z_1/z_2 =1.64$, see Eq.~(\ref{eq:asymp1}) and the comments
immediately below it. Given that the latter
result is expected to hold for $z \gg 1$, one may infer that the small discrepancy found
is due to $z_1$, $z_2$ not being large enough.

Finally we remark that the pure-ferromagnet limit is somewhat subtle, because in order 
to reach 
the stationary regime of the F-P equation where the $N$-independent PDF forms hold, 
the condition $N \gtrsim \lambda$ must be obeyed. However, as $p \to 1$, and for the low
energies where scaling is valid, $\lambda$ diverges as one approaches the 
pure-system magnon band (located at $0 \leq \omega < 2$, in the current 
units). For example, at $p=0.95$ and $\omega=0.015$, one gets $\lambda 
\approx 2500$. 

\subsection{Non-stationary regime for $p \neq \frac{1}{2}$}
\label{sec:3e}

Still for the magnetized case $p \neq 1/2$, one can learn more by considering selected 
aspects of the non-stationary regime. Going back to Eq.~(\ref{eq:fp-mag}), one sees 
that, in the limit $p-1/2 \gg \omega^{1/3}$, i.e., $z \gg 1$, the diffusive term becomes 
small and this regime is associated with just streaming in leading order. The
non-stationary equation to be solved is thus:
\begin{equation}
\frac{\partial P(s,\ell)}{\partial \ell}= 
\frac{\partial}{\partial s}\left((s^2+\omega\,{\cal M}_1)\,
P(s,\ell)\right)\ .
\label{eq:fp-mag-ns}
\end{equation}
With $\beta\equiv[\omega\,(p-1/2)]^{1/2}$, ${\tilde t}\equiv \ell\beta$,
 ${\tilde y}\equiv s/\beta$, and defining $Q({\tilde t},{\tilde y})$ such that
$ Q({\tilde t},{\tilde y})\,d{\tilde y}=P(s,\ell)\,ds$, Eq.~(\ref{eq:fp-mag-ns})
turns into:
\begin{equation}
\frac{\partial Q}{\partial {\tilde t}}=\frac{\partial }{\partial {\tilde y}}
\left({\tilde y}^2+1\right) Q\ ,
\label{eq:Q1}
\end{equation}
whose general solution is:
\begin{equation}
Q({\tilde t},{\tilde y})= \frac{1}{1+{\tilde y}^2}\,f \left({\tilde t} +\tan^{-1}
{\tilde y}\right)\ .
\label{eq:Q2}
\end{equation}
In order to have the condition $p-1/2 \gg \omega^{1/3}$ fulfilled to a good extent,
so that the scaling behavior predicted by Eq.~(\ref{eq:Q2}) could be 
unequivocally demonstrated, we used $p=0.6$, $\omega =10^{-6}$. For such a choice, the
localization length is $\lambda \gtrsim 4\times 10^4$. This allowed us to take values
of $N$ equal to several hundred lattice spacings, which are both much larger than unity
(so discrete-lattice effects are negligible), and still much shorter than $\lambda$
(thus guaranteeing non-stationarity by a broad margin).   
\begin{figure}
{\centering \resizebox*{3.3in}{!}{\includegraphics*{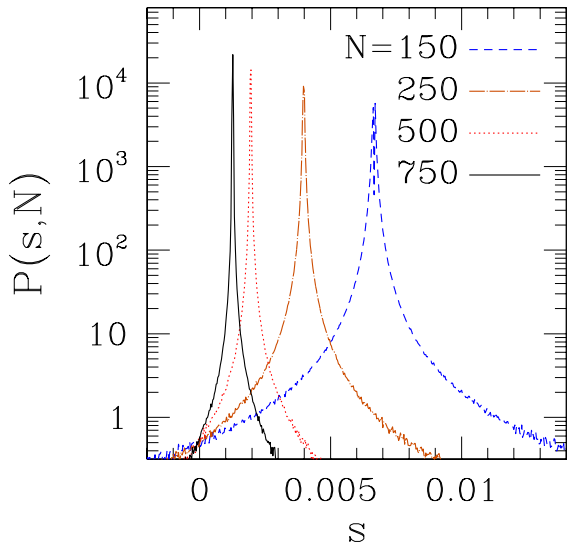}}}
\caption{(Color online) Central sections of normalized histograms of occurrence of 
instantaneous LLE  $s$, from  $N_s=10^7$ samples, for $\omega=10^{-6}$, $p=0.60$ 
and $N$ as indicated. 
} 
\label{fig:nsraw}
\end{figure}
Figure~\ref{fig:nsraw} shows the region close to the central peaks of the PDFs 
(where scaling is expected to hold), for $N=150$, $250$, $500$ and $750$.  In 
Fig.~\ref{fig:nsscale} the same data are shown in a scaling plot, as suggested by
Eq.~(\ref{eq:Q2}). Note that the number of samples $N_s$ is one order of magnitude 
larger than, e.g., that used in earlier sections of this work. This was necessary, in 
view of the relatively wide scatter of the PDFs: typically, they displayed almost flat 
tails running out to $|s| \gtrsim 0.07$, so the relevant data as far as scaling is 
concerned were a small subset of the total gathered (compare the horizontal scale in
Fig.~\ref{fig:nsraw}). Even so, one can see, close to the bottom of the scaling curve
in Fig.~\ref{fig:nsscale}, that a non-negligible degree of fluctuation-induced spread 
still remains.  Nevertheless, overall agreement with the scaling predictions of
Eq.~(\ref{eq:Q2}) is remarkable.

\begin{figure}
{\centering \resizebox*{3.3in}{!}{\includegraphics*{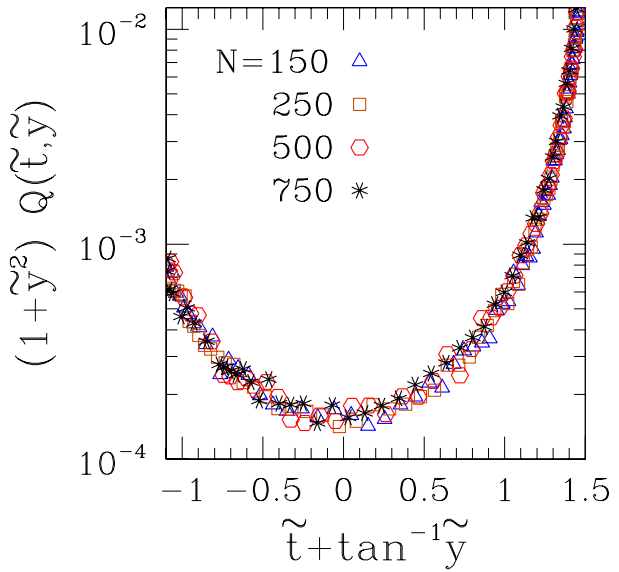}}}
\caption{(Color online) Scaling plot of data displayed in 
Figure~{\protect{\ref{fig:nsraw}}}, as suggested by
Eq.~({\protect{\ref{eq:Q2}}}).
} 
\label{fig:nsscale}
\end{figure}

\section{Discussion and Conclusions} 
\label{sec:conc}
We have investigated scaling properties of the probability distribution functions 
(PDF) of Lyapunov exponents for the  one-dimensional Heisenberg-Mattis spin glass. 

In Section~\ref{sec:3a} we showed that, for a given energy $\omega$ (small enough, such
that the localization length $\lambda(\omega)$ is sufficiently large for 
scaling concepts to apply), the PDF of the local Lyapunov exponent (LLE) 
$\gamma(N,\omega)$, takes on a shape which, for $N \gtrsim \lambda(\omega)$, is 
increasingly close to a Gaussian. In the limit $N \to \infty$, the PDFs turn into
$\delta$ functions, in conformity with the central limit theorem. Thus, such aggregate 
effects reflect only general statistical properties of fluctuations in systems with many 
(almost-) independent degrees of freedom. The 
connection with the underlying physical problem is traced exclusively through the 
dependence of the numerical value of the (asymptotic) Lyapunov exponent with energy (and 
ferromagnetic bond concentration; though this latter aspect was not exploited here, it 
has been investigated before~\cite{bh89,ew92,ah93}).   

On the other hand, the PDFs for the instantaneous Lyapunov exponents, as defined via
Eq.~(\ref{eq:instdef}), display the following properties of interest, exhibited
in Sections~\ref{sec:3b}, \ref{sec:3c}, and~\ref{sec:3d}: (i) for
$N \gtrsim \lambda(\omega)$ they approach a fixed, non-trivial shape with nonvanishing 
width; (ii) such specific shape can be predicted through a connection with the 
stationary  state of a Fokker-Planck (F-P) equation; and (iii) the shape is a 
``fingerprint" shared by all the systems in the universality class in which the 
statistical properties of the system lie . Furthermore, the F-P equation 
providing the link can be set up directly from, and closely reflects, the
physical features of the system under investigation [ see especially 
Eqs.~(\ref{eq:cont1})--(\ref{eq:fp1})].

The applicability of a description via an F-P equation goes beyond the stationary state, 
as
shown in Section~\ref{sec:3e}. There, it is shown that the scaling of non-stationary PDFs
in a specific regime (in which diffusion effects are expected to be negligible)   
closely follows predictions drawn from the corresponding form of the F-P equation.

In the present work we have demonstrated that
an F-P approach can be successively applied to both stationary and non-stationary
properties of the PDF of instantaneous Lyapunov exponents for Heisenberg-Mattis spin 
glasses. 

As a final remark, we believe that the scaling properties of
the PDF of Lyapunov exponents described here are identical to those
pertinent to the zero-temperature random-bond ($\pm 1$) classical
Heisenberg chain.
This comes via the identification of the scaling properties of
the low-energy excitations of the HSMG chain with those of
the classical Heisenberg model in 1d \cite{sp88,ps88,ew92,bh89}, plus the
general applicability of the F-P equation in the continuum~\cite{fp1}.

It is expected that treatments along these lines can be devised, with
similar degree of success, for other physical systems in whose description Lyapunov 
exponents play a prominent role.

\begin{acknowledgments}
S.L.A.d.Q. thanks the Rudolf Peierls Centre for Theoretical Physics,
Oxford, where this work was initiated, for the hospitality,
and CNPq and Instituto do Mil\^enio de Nanoci\^encias--CNPq for 
funding his visit. R.B.S. wishes
to thank Dr. Alexandre Lefevre for discussions of fundamental aspects
concerning the Fokker-Planck description used here.
The research of S.L.A.d.Q. was partially supported by
the Brazilian agencies CNPq (Grant No. 30.0003/2003-0), FAPERJ (Grant
No. E26--152.195/2002), and Instituto do Mil\^enio de
Nanoci\^encias--CNPq.
R.B.S. acknowledges partial support from EPSRC Oxford Condensed Matter
Theory Programme Grant EP/D050952/1.
\end{acknowledgments}

\end{document}